# End-to-End V2X Latency Modeling and Analysis in 5G Networks

B. Coll-Perales, *Member, IEEE,* M.C. Lucas-Estañ, *Member, IEEE,* T. Shimizu, *Member, IEEE,* J. Gozalvez, *Senior Member, IEEE,* T. Higuchi, *Member, IEEE,* S. Avedisov, *Member, IEEE,* O. Altintas, *Member, IEEE,* M. Sepulcre, *Senior Member, IEEE*

*Abstract*—5G networks provide higher flexibility and improved performance compared to previous cellular technologies. This has raised expectations on the possibility to support advanced Vehicle to Everything (V2X) services using the cellular network via Vehicle-to-Network (V2N) and Vehicle-to-Network-to-Vehicle (V2N2V) connections. Replacing direct Vehicle-to-Vehicle (V2V) connections by V2N and V2N2V communications to support critical V2X services requires low V2N and V2N2V latencies. It is then necessary to quantify the latency values that V2N and V2N2V connections can achieve over 5G end-to-end (E2E) connections, but these depend on the particular 5G network deployments and configurations and to date the number of related studies is limited. Most existing studies focus on evaluating the performance and feasibility of the 5G radio access network to support advanced V2X services, or consider dedicated 5G pilot deployments under controlled conditions. The network deployment and dimensioning also have a strong impact on the E2E performance, and hence on the suitability of V2N/V2N2V connections to support advanced V2X services with low latencies. This paper progresses the state-of-the-art by introducing a novel E2E latency model to quantify the latency of 5G V2N/V2N2V communications. The model includes the latency introduced at the radio, transport, core, Internet, peering points and application server (AS) for single mobile network operator (MNO) and multi-MNO scenarios. This paper estimates the E2E latency for a large variety of possible 5G network deployments that are being discussed or envisioned to support V2X services. This includes the possibility to deploy the V2X AS from the edge of the network to the cloud. The model is utilized to analyze the impact of different 5G network deployments and configurations on the E2E latency. The analysis helps identify which 5G network deployments and configurations are more suitable to meet V2X latency requirements, using the cooperative lane change as a case study. The conducted analysis highlights the challenge for centralized network deployments that locate the V2X AS at the cloud to meet the latency requirements of advanced V2X services. Locating the V2X AS closer to the cell edge reduces the latency. However, it requires a higher number of ASs and also a careful dimensioning of the network and its configuration to ensure sufficient network and AS resources are dedicated to serve the V2X traffic.

## I. INTRODUCTION

5G networks provide unprecedented levels of flexibility and adaptability that are necessary to support critical applications with stringent requirements in key vertical markets, including automotive and mobility. The support for these verticals is expected to grow with the use of intelligent reflecting surfaces in future networks and the possibility to offload processing tasks from vehicles to the infrastructure [1]-[3]. Vehicle to Everything (V2X) services have been traditionally envisioned with direct or sidelink connections between vehicles (Vehicle-to-Vehicle, V2V) or between vehicles and infrastructure nodes (Vehicle-to-Infrastructure, V2I) for the most critical services. Connections through the cellular network (Vehicle-to-Network, V2N) are usually envisioned for infotainment, comfort, or traffic management services. However, the capabilities of 5G have raised questions and expectations on the possibility to support critical services using the cellular network and V2N2V (Vehicle-to-Network-to-Vehicle) connections rather than direct or sidelink V2V communications[1]. For this to happen, it is necessary to ensure that 5G network deployments and configurations can guarantee the end-to-end (E2E) performance demanded by V2X critical services. This includes satisfying the stringent latency requirements demanded by safety critical V2X services.

Most of the studies that analyze the latency of 5G networks when supporting V2X services focus on the latency introduced by the 5G radio access network [4]-[6]. For example, [6] reports uplink (UL) and downlink (DL) radio latencies from 1 ms to 5 ms depending on the 5G New Radio (NR) configuration. The latency introduced by the transport and core networks is usually modeled using fixed values. For example, [7] considers that the one-way core network (CN) latency is 200 μs or 100 μs for non-standalone (NSA) and standalone (SA) networks, respectively, and that the latency introduced in the link between the CN and the location of the V2X Application Server (AS) is 5.4 μs. The CN latency is estimated in [7] considering the processing delay introduced by the CN nodes, but does not take into other effects like the propagation or queuing delays. In addition, [7] does not model or estimate the latencies introduced at the transport network (TN). The studies in [8]-[10] model the 5G network processing latency with a single value of 20 ms. The network processing latency is the time between the moment a base station receives a packet from a vehicle in the UL, and the moment at which the base station transmits it in the DL.

UMH work was supported in part by MCIN/AEI/10.13039/501100011033 (grants IJC2018-036862-I, PID2020-115576RB-I00) and by Generalitat Valenciana (GV/2021/044, CIGE/2021/096).

B. Coll-Perales, M.C. Lucas-Estañ, J. Gozalvez and M. Sepulcre are with Universidad Miguel Hernández de Elche, Spain. Contact emails: {bcoll, m.lucas, j.gozalvez, msepulcre}@umh.es.

T. Shimizu, T. Higuchi, S. Avedisov, and O. Altintas are with InfoTech Labs, Toyota Motor North America R&D, Mountain View, CA, U.S.A. Contact emails: {takayuki.shimizu, takamasa.higuchi, sergei.avedisov, onur.altintas}@toyota.com

[1] V2N and V2N2V are also referred to as V2C (Vehicle-to-Cloud) and V2C2V to reflect the processing of V2X packets at edge or center clouds.

TABLE I. SUMMARY OF PREVIOUS RELATED WORKS

| Reference | 5G latency component(s) | Values | Conditions/Comments |
|---|---|---|---|
| [6] | Radio Access Network | [1, 5] ms | UL + DL latency values depending on the 5G NR configuration |
| [7] | CN | 200 μs (NSA) or 100us (SA) | 1-way uplink latency. Does not consider queuing delays, impact of load, and varying propagation delays depending on the network deployment. Additional 5.4 μs-delay considered in the link between the CN and V2X AS. |
| [8]-[10] | TN + CN | 20 ms | Does not consider queuing delays, impact of load, and varying propagation delays depending on the network deployment. |
| [11] | CN | > 2 ms | Considers queuing, processing, and transmission delays. The total CN value depends on the network load and the number of intermediate nodes. It does not account for propagation delays. |
| [12] [13] [15] | MEC-based end-to-end V2N2V | 7.8 ms [15, 20] ms | Mean value measured in field tests with the V2X AS located at the MEC. |
| [12] [13] [14] | Cloud-based end-to-end V2N2V | [53.8, 150] ms | Mean value measured in field tests with the V2X AS located on the cloud. Includes 10-12ms Internet latency. Latency increases to 201 ms for scenarios where vehicles are supported by different operators. |

Existing studies provide first insights into the 5G latency for supporting V2X services. However, the use of fixed values to model the latency does not consider the impact of 5G network deployments and configurations, and varying network traffic loads on the latency. An important contribution is reported in [11] where authors introduce a latency model for the 5G core network that considers the packet queuing, processing, and transmission delays. [11] shows that the total CN delay depends on the network load and the number of intermediate switching nodes processing the data. However, [11] only focuses on the core network and an E2E perspective is necessary for a comprehensive evaluation of the capabilities of 5G to support V2X services via V2N and V2N2V links. This is a challenge tasks since the 5G E2E latency depends on the radio access network configuration, the transport and core networks' deployments, the location at which the V2X ASs are deployed, the deployment of Mobile/Multi-Access Edge Computing (MEC) nodes, as well as the use of peering points between networks and Internet connections when necessary. To the best of the authors' knowledge, there are only few studies that have analyzed the E2E performance of 5G V2N and V2N2V communications, and most of them are based on field trials conducted in dedicated pilot deployments. This is for example the case of [12] that reports mean E2E latency values for MEC-based network deployments of 17.8 ms. Out of this, [12] shows that the processing of the data at the application server takes approximately 10 ms, which results in around 7.8 ms for the 5G E2E latency. This latency value is in line with some preparatory trials reported in [12] that show an 8.2 ms E2E latency using a ping application in a MEC-based network deployment. The reported mean E2E latency values increase to 53.8 ms when the V2X AS is located at a central cloud reached via the Internet. The average latency introduced by the Internet connection is between 10 ms to 12 ms. Field tests conducted in [13] report 5G E2E latency ranging from 15 ms to 20 ms for MEC-based deployments. The results reported in [14] show 5G E2E latency to an AS located at the cloud. Preliminary field tests over 5G non-standalone deployments show E2E latencies ranging from 23 ms to 201 ms in [13] for scenarios where vehicles are supported by different network operators. Information about the network deployments and configurations are limited in [13] and [14]. In [15], the targeted 5G V2N2V E2E latency is below 30 ms for network deployments supported by MEC in a corridor between Germany, Austria, and Italy. The obtained results are not reported yet though. The E2E latency values reported in [8] for 5G V2N2V E2E connections range from 23.9 ms to 40.8 ms and from 23.9 ms to 79.5 ms when the DL transmissions use unicast or broadcast mode, respectively. The variations in the E2E latency values reported in [8] depend on the configured slot duration in the 5G radio interface, and a fixed 20 ms latency is considered for the 5G network processing latency. [9] uses the results reported in [8] to model the 5G V2N2V latency using a uniform distribution between 60 ms and 130 ms independent of the particular network deployment. Table I summarizes the analysis of the 5G latency results reported in previous works.

The motivation of this work comes from the increasing expectations put in 5G networks to support critical V2X services using the cellular network and V2N2V connections rather than direct or sidelink V2V communications. The flexibility introduced in 5G increases the number of possible deployments and configurations. This makes it challenging to develop accurate simulation and testbeds to evaluate the capabilities of 5G to support V2X services via V2N and V2N2V links. The existing work in the literature is trying to derive conclusions on the capability of 5G to support the requirements of V2X services through: 1) V2N2V connections based only on specific 5G pilot deployments under controlled environments; and 2) numerical results that only account the performance of the 5G radio access network or consider fixed network processing latencies.

In this context, we aim to complement existing studies and progresses the state-of-the-art by proposing a novel 5G E2E latency model. We believe that the proposed model is a unique contribution to the community that will allow researchers to estimate 5G E2E latency values without having to deploy, or do measurements, over a real and complete 5G network, which is something accessible to just a few colleagues. The E2E latency model accounts for the latency contributions of the radio network, TN, CN, possible deployment of MEC nodes, peering points among networks, connections to the Internet, and processing capabilities of the AS that hosts the V2X applications. This new model significantly expands the authors' previous contribution in [16] that only modeled the latency at the radio network level. The model presented in this study represents, to the authors' knowledge, the first E2E 5G latency model that can quantify the end-to-end latency experienced by V2N and V2N2V communications over different 5G network deployments. It hence represents a valuable tool to the community to identify which services, and under which network deployments and configurations, can be supported with 5G V2N or V2N2V communications.

In summary, the contributions of this study are:

- We present a novel 5G E2E latency model for V2N and V2N2V communications. The model includes the latency introduced at the radio, transport and core networks, Internet, peering points and the V2X AS.
- We derive expressions of the 5G E2E latency for centralized 5G network deployments where the V2X AS is hosted at the cloud, and for 5G network deployments where the V2X AS is located at the edge, transport, or core networks.
- We derive expressions of the 5G E2E latency in single mobile network operator (MNO) and multi-MNO scenarios.
- We quantify and analyze the 5G E2E latency performance for V2N and V2NV2N communications under various 5G network deployments and configurations.
- We identify which 5G network deployments and configurations can better meet the latency and reliability requirements of V2X services for connected and automated driving with low or high level of automation. To this aim, we use the cooperative lane change service as a case study.

## II. OVERVIEW OF 5G

5G standards define a flexible 5G CN along with an advanced radio access network (RAN) that includes a NR interface. Both RAN and CN are highly flexible to support a wide range of service requirements and network deployments. While broadcast and multicast communications are important for V2X services, they are not supported in 3GPP Release 16 but are under study in Release 17.

### A. 5G Architecture

Fig. 1 represents the 5G architecture [17] . The Uu interface between the 5G base station (gNB) and the vehicle is used for data plane and control plane communications. The RAN is layered into a radio network layer and a TN layer. The TN interconnects the RAN and CN through a set of multiplexing nodes that are deployed following a hierarchical architecture that form the so-called access and aggregation domains [18]. At the 5G CN, hardware network elements used in traditional cellular systems are replaced by network functions that can be virtualized (e.g., Virtual Network Functions, VNF) and implemented in software. This facilitates the separation of the control and user plane's network functions and the realization of flexible deployments. The network functions are connected using interfaces or reference points, referred to as Nx in Fig. 1. Connectivity services at the 5G CN are supported by one or more User Plane Functions (UPFs) that route the traffic through the CN to the Data Network (DN). UPFs are logically part of the CN, but 5G allows to install them at any location from the gNB to the CN. The E2E data plane connectivity between the vehicle and the DN is referred to as packet data unit session, and the UPF that provides access to the DN is called packet data unit session anchor (PSA UPF). The flexibility introduced by virtualized core networks support a large diversity of 5G network deployments, and it is important to understand and quantify their impact on the latency of 5G E2E connections.

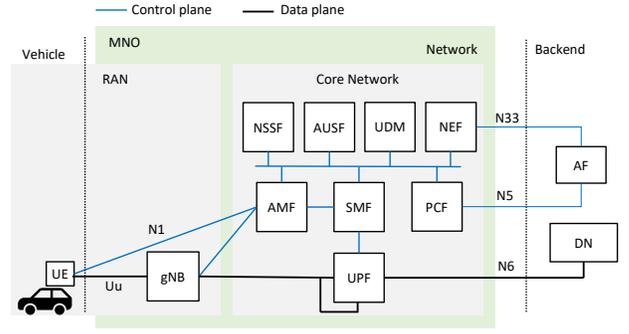

Fig. 1. 5G System architecture [17].

### B. MEC

5G supports the integration of MEC and network slicing to support services with distinct and stringent communication requirements. Network slicing creates different logical partitions or slices of the common physical network infrastructure. Each network slice can be designed and configured to support specific services with different QoS requirements while guaranteeing their isolation.

MEC brings computing and storage resources, as well as management functions, to the edge of the network to reduce the latency. Keeping the information and processing closer to the users also reduces the traffic load at the CN and facilitates the scalability of the network. The integration of MEC in the 5G network is done through the N6 interface of the UPF (Fig. 1). The data is routed in the 5G network through a UPF (or a chain of UPFs) towards/from the MEC system. The MEC system is made of MEC hosts and the MEC management functions that orchestrate its operation [19]. The flexibility in locating the UPF makes it possible to have different MEC deployments. This is possible thanks to the 5G's local routing and traffic steering functionalities that identify the appropriate PSA UPF that provides access to the MEC through the optimal path.

### C. 5G New Radio

5G NR introduces the use of flexible numerologies. Each numerology defines a different subcarrier spacing (SCS) in the frequency domain and slot duration in the time domain (Table II). A slot consists of 14 or 12 OFDM (Orthogonal Frequency-Division Multiplexing) symbols when normal and extended cyclic prefix is used, respectively. 5G NR supports time and frequency division duplex. The definition of the frame in time division duplex mode is highly flexible but it has synchronization challenges. With frequency division duplex, UL and DL resources are always available. 5G NR allows transmissions using the full-slot (i.e., 14 or 12 OFDM symbols in a slot), but also mini-slots with limited OFDM symbols [20].

TABLE II. NUMEROLOGIES AND SLOT DEFINITION [20]

| Numerology | SCS [kHz] | Cyclic prefix | OFDM symbols per slot | Slot time duration [ms] |
|---|---|---|---|---|
| 0 | 15 | Normal | 14 | 1 |
| 1 | 30 | Normal | 14 | 0.5 |
| 2 | 60 | Normal, extended | 14/12 | 0.25 |
| 3 | 120 | Normal | 14 | 0.125 |
| 4 | 240 | Normal | 14 | 0.0625 |

5G NR also defines three different MCS (Modulation and Coding Scheme) index tables in [21] for data transmission both in UL and DL. Table 1 and Table 2 in [21] provide high spectrum efficiency, and a target BLER (Block Error Ratio) of 0.1. Table 3 in [21] provides a lower spectrum efficiency and target BLER of 10$^{-5}$.

## III. 5G NETWORK DEPLOYMENTS

5G network deployments have important performance and technical implications for V2X service provisioning. The implementation of V2X ASs at the edge reduces the latency but can challenge service continuity due to possible frequent handovers between MEC hosts as vehicles move [22]. On the other hand, centralized 5G network deployments where the AS is located at the cloud may facilitate service continuity as such deployments cover larger areas at the expense of higher latencies. We then analyze the main possible 5G network deployments considered to support V2X services. We follow for al deployments the reference network architecture defined by the 5G Automotive Association in [22]. We also consider that the RAN and the 5G CN are interconnected using the hierarchical TN proposed by ITU-T (International Telecommunication Union Telecommunication Standardization Sector). This TN integrates 3 multiplexing nodes M1, M2 and M3 (see Fig. 2) [18]. The 5G network deployments analyzed in this study are [22]:

- *Centralized.* This deployment considers that the V2X AS is located at the cloud outside the mobile network domain (Fig. 2.a). This deployment can benefit from powerful computing and storage resources available on the cloud and facilitate the deployment of V2X services that can reside, for example, at an original equipment manufacturer cloud. However, this deployment increases the latency as V2X packets need to traverse the entire network, and Internet access is required to reach the V2X AS.
- *Core Central Office* (MEC@CN). In this deployment, a MEC is deployed at the CN and hosts the V2X AS (Fig. 2.b). It requires routing V2X packets through the TN, and it uses a UPF of the 5G CN to steer the traffic towards the MEC.
- *Central Office* (MEC@*M1*). In this deployment the MEC is collocated with the multiplexing node M1 (Fig. 2.c). A UPF is implemented locally in the same location of M1 to steer the traffic towards the V2X AS running in the MEC.
- *Cell site* (MEC@*gNB*). This deployment considers that the MEC and the local UPF are collocated at the gNB (Fig. 2.d).

For simplicity of discussion, Fig. 2 assumes that a single MNO (Mobile Network Operator) is serving all vehicles. If vehicles are supported by different MNOs, these MNOs interact through interconnection links called peering points. These interconnections are referred to as remote (public) peering points when they are established through the public Internet at Internet exchange points, or as local (private) when they use direct links between the MNOs CN's UPF nodes.

## IV. 5G END-TO-END LATENCY MODELING

This section presents the model derived to estimate the end-to-end latency ($l_{E2E}$) experienced in 5G V2N and V2N2V communications. For V2N communications, the model considers the latency experienced by a packet transmitted by a vehicle to a V2X AS in the UL path (or reverse path in DL). UL and DL paths are both considered for V2N2V communications.

We derive the general expression of the E2E latency $l_{E2E}$ for a Centralized network deployment. Fig. 3 illustrates the communication path between different vehicles and the V2X AS for a Centralized deployment. Fig. 3 shows the case when the vehicles (UE1 and UE2) are supported by the same MNO, and the case when vehicles (UE1 and UE3) are supported by different MNOs. When vehicles are supported by the same MNO, $l_{E2E}$ can be expressed as the sum of the latency experienced at the RAN ($l_{RAN}$), CN ($l_{CN}$), in the communication link between the PSA UPF of the CN and the V2X AS ($l_{UPF-AS}$), and the latency introduced by the V2X AS ($l_{AS}$). $l_{RAN}$ is expressed as the sum of $l_{radio}$ and $l_{TN}$ that represent the latency of packets in the radio and transport networks, respectively. Then, $l_{E2E}$ can be expressed as:

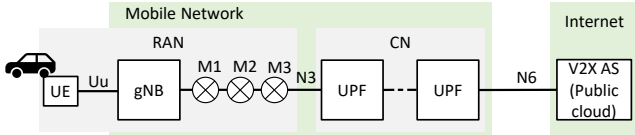
a) Centralized: V2X AS located in the cloud.

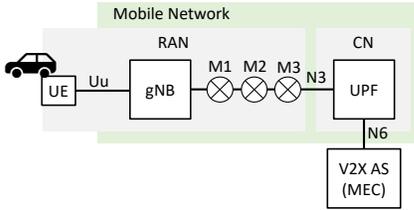
b) Core Central Office: MEC and V2X AS deployed at the CN.

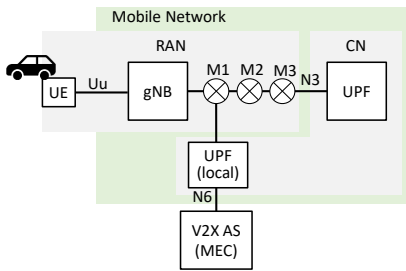
c) Central Office: MEC and V2X AS located at the M1 node.

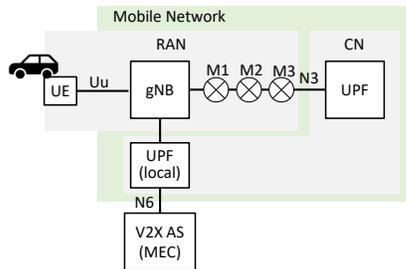
d) Cell site: MEC and V2X AS located at the gNB.

Fig. 2. 5G V2X deployment scenarios.

$$l_{E2E} = l_{RAN}+l_{CN}+l_{UPF-AS}+l_{AS} = l_{radio}+l_{TN}+l_{CN}+l_{UPF-AS}+l_{AS} \quad (1)$$

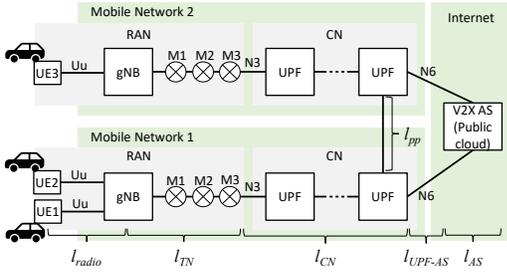

Fig. 3. Latency components affecting V2N2V E2E latency.

When vehicles are supported by different MNOs, $l_{E2E}$ also includes the latency $l_{pp}$ experienced in the peering point between MNOs (Fig. 3):

$$l_{E2E} = l_{radio} + l_{TN} + l_{CN} + l_{UPF-AS} + l_{AS} + l_{pp}. \quad (2)$$

For each latency component of $l_{E2E}$ in (2), we calculate the latency $l_x^{UL}$ experienced in UL and the latency $l_x^{DL}$ experienced in DL, $x \in \{radio, TN, CN, UPF\text{-}AS, PP\}$. Then, the round-trip latency $l_x$ in a network component $x$ is computed as $l_x^{DL}+l_x^{UL}$. The average value of the latency is obtained as $\overline{l_x} = \overline{l_x^{UL}} + \overline{l_x^{DL}}$, and its cumulative distribution function (CDF) as $\mathcal{L}_x(t) = \mathcal{L}_x^{UL}(t) * \mathcal{L}_x^{DL}(t)$.

Eq. (2) can be used for all 5G network deployments described in Section III, but with different expressions and values for $l_{TN}$, $l_{CN}$, $l_{UPF-AS}$, $l_{AS}$, and $l_{pp}$ since they depend on the particular 5G network deployment. This is not the case of $l_{radio}$ that only depends on the configuration of the radio network. In the following sections we derive the expressions of $l_{TN}$, $l_{CN}$, $l_{UPF-AS}$, $l_{AS}$, and $l_{pp}$ for all 5G network deployments. To model $l_{radio}$ we use our results from [16] where we considered many different configuration options available for 5G NR at the radio network (e.g., numerology, scheduling, and retransmissions). We summarize the modelling of $l_{radio}$ from [16] in the next section.

## V. 5G RADIO LATENCY

The authors have estimated the radio latency $l_{radio}$ in [16]. $l_{radio}$ is the sum of the latency $l_{radio}^{UL}$ experienced in the transmission of a packet from a vehicle to the gNB, and the latency $l_{radio}^{DL}$ experienced in the related downlink transmissions from the gNB to all neighboring vehicles interested to communicate with the vehicle that transmitted the packet in the UL. [16] estimates $l_{radio}$ considering scenarios where DL transmissions are conducted using multiple unicast transmissions or broadcast transmissions.

$l_{radio}$ has been derived for different 5G NR configurations characterized by the numerology or SCS, slot duration, retransmission, and scheduling schemes, as well as system parameters such as the allocated bandwidth, the characteristics of the data traffic, and the density and the distribution of vehicles. The UL ($l_{radio}^{UL}$) and DL ($l_{radio}^{DL}$) latencies account for the latency $T_i^X$ in the initial transmission of a packet, and the latency $n \cdot T_r^X$ introduced by the transmission of $n$ additional repetitions of the packet or $n$ potential HARQ retransmissions. We note that in this paper we use the superscript $X$ to indicate UL or DL directions. $T_i^X$ is estimated considering the latency introduced by the scheduling process, and the time needed to transmit and receive the initial packet. The radio transmission time $t_{tt}$ is equal to the slot time duration and depends on the numerology (Table II). $T_r^X$ accounts for the signaling exchanged to request the repetition/retransmission of a packet, and the latency introduced by the transmission of this packet.

## VI. TRANSPORT NETWORK LATENCY

The transport network (TN) latency $l_{TN}$ is the sum of $l_{TN}^{UL}$ and $l_{TN}^{DL}$. $l_{TN}^{UL}$ accounts for the processing of packets at the gNB and their transmission time over the TN towards the V2X AS. $l_{TN}^{DL}$ accounts for the transmission time of packets over the TN towards the gNB and their processing at the gNB. $l_{TN}^X$ can be expressed as:

$$l_{TN}^X = t_{TN,prop}^X + t_{TN,transit}^X, \quad (3)$$

where $t_{TN,prop}^X$ is the propagation delay or time it takes a packet to travel through the links that interconnect the nodes of the TN. $t_{TN,transit}^X$ is the transit delay and represents the time that packets spend in the TN's node(s). We consider in (3) that packets do not require retransmissions since this is the case in most network links in practice [23][24]. $t_{TN,prop}^X$ depends on the total distance $d_{TN}$ that packets travel through the TN, and the propagation speed $v$ in the TN medium:

$$t_{TN,prop}^X = \frac{d_{TN}}{v}. \quad (4)$$

$t_{TN,transit}^X$ accounts for delays associated to the TN nodes' reception and transmission processing. It includes:

- Processing delays ($t_p$) between the time a packet is correctly received, and it is assigned to an outgoing link queue for transmission,
- Queuing delays ($t_q$) between the time a packet is assigned to a queue for transmission and the time it starts being transmitted,
- Transmission time delays ($t_{tt}$).

To estimate $t_{TN,transit}^X$, this study models the TN as a queuing system. Following [25], we model each node of the TN as an M/M/1 queue[2]. Each node of the TN is represented as a single server that processes packets that arrive following a Poisson process with average rate $\lambda$. The node serves or dispatches the received packets at an average service rate $\mu$ that follows an exponential distribution. The utilization of a node is calculated as $\rho = \lambda/\mu$. M/M/1 queues consider infinite buffer capacity. However, we consider that the TN nodes are not stable if $\rho > 1$ (i.e., $\lambda > \mu$). This is the case because the buffer backlog would increase to infinity and there would be packets that never depart the node in a finite amount of time. Using $\lambda$ and $\mu$, the average transit delay when packets pass through a set of $n$ nodes in the TN can be expressed following Jackson's theorem as [24]:

$$\overline{t_{TN,transit\_n}^X} = n(t_p + t_q + t_{tt}) = n \cdot t_p + \sum_i \frac{1}{\mu_i^X - \lambda_i^X}, \quad (5)$$

---

[2] This queuing system fits adequately with the most general case in which packets arrive to the TN following a Poisson process. Other traffic patterns (e.g., bursty or event.-based) might require a different modeling that is out of the scope of this paper.

where $\lambda_i^X$ and $\mu_i^X$ are the arrival rate of traffic and the service rate of the node $i$, $i=1,\ldots, n$. Following (5), it is necessary to calculate $\lambda_i^X$ and $\mu_i^X$ for each TN node $i$ to estimate $\overline{t_{TN,transit\_n}^X}$. For that, we can then use the properties of Poisson processes. If we consider a hierarchical architecture for the TN, it can be established that a TN node $i$ aggregates traffic from a set of $g$ TN nodes in the UL. In this case, $\lambda_i^{UL}$ can be computed as $\sum_{k=1}^{g} \lambda_k^{UL}$. In the DL, the TN node $i$ splits traffic among the $g$ TN nodes. The arrival rate of traffic that the node $i$ has to dispatch through each of the $g$ links is determined by a Poisson process with average rate $\lambda_i^{DL} = p_k \cdot \lambda_{i,\,total}^{DL}$. $p_k$ is the portion of the overall arrival rate of traffic $\lambda_{i,\,total}^{DL}$ that is addressed to the output link $k$, $k=1,\ldots, g$. All packets that arrive to the TN node $i$ have to be dispatched, so the condition $\sum_{k=1}^{g} p_k = 1$ must be satisfied. The service rate of a TN node $i$ in the UL is computed as:

$$\mu_i^{UL} = \frac{\alpha_{i\text{-}(i+1)} \cdot C_{i\text{-}(i+1)}}{B}, \quad (6)$$

where $B$ is the size (in bits) of the packets that are transmitted over the links of the TN, $C_{i\text{-}(i+1)}$ (in bits/s) represents the capacity of the link that connects node $i$ and the next node $i+1$ in the UL path, and $\alpha_{i\text{-}(i+1)}$ is the fraction of this link capacity that is allocated in the UL to support the traffic of a specific service. The service rate of the node $i+1$ in the DL $\mu_{i+1}^{DL}$ is computed as in (6) but using $\alpha_{(i+1)\text{-}i}$ to show the opposite direction of the link. Different values of $\alpha$ may be allocated to support UL and DL traffic over the same link between nodes $i$ and $i+1$, subject to a sum of both values lower than 1.

Equations (4) and (5) are used to derive $\overline{l_{TN}}$. The cumulative distribution of the TN latency can then be estimated as:

$$\mathcal{L}_{TN}^X(t) = \begin{cases} 1 - e^{-\delta(t - t_{TN,prop}^X)} & \text{if } t \geq t_{TN,prop}^X, \\ 0 & \text{otherwise} \end{cases} \quad (7)$$

with $\delta = 1/\overline{t_{TN,transit}^X}$.

Equations (3) to (7) are the general expressions for the TN latency and apply to all 5G network deployments defined. However, the deployments differ in the values of the $t_{TN,prop}^X$ and $t_{TN,transit}^X$ variables that are computed next for each deployment.

*A. MEC@gNB*

This deployment considers that the gNB and MEC hosting the V2X AS are collocated. Then, $d_{TN}$ can be approximated to 0 m, and hence $t_{TN,prop\_gNB}^X = 0$ s.

Packets received at the gNB are processed and transmitted towards the V2X AS via the UPF in the UL. In the DL, packets coming from the UPF are processed at the gNB and then transmitted via the radio interface. Then, following (5), the transit delay in the UL and DL can be computed as:

$$\overline{t_{TN,transit\_gNB}^{UL}} = t_p + \frac{1}{\mu_{gNB}^{UL} - \lambda_{gNB}^{UL}} \quad (8)$$

$$\overline{t_{TN,transit\_gNB}^{DL}} = t_p \quad (9)$$

The arrival rate of traffic to the gNB in the UL is then:

$$\lambda_{gNB}^{UL} = N_{UE} \cdot T_p \cdot (1 - P_{rel}), \quad (10)$$

where $N_{UE}$ is the number of UEs or vehicles under the coverage of the gNB. Without loss of generality, (10) considers that all vehicles under the coverage of the gNB generate packets of $B$ bits at an average rate of $T_p$ packets/s. $P_{rel}$ is the reliability in the UL radio interface. $P_{rel}$ can be approximated to the BLER when vehicles only perform one transmission per packet, or to $1 - \text{BLER}^{n+1}$ when vehicles perform $n$ additional transmissions with repetitions/retransmissions of the original packet. Following (6), the service rate of the gNB is expressed as $\mu_{gNB}^{UL} = \frac{\alpha_{gNB\text{-}UPF} \cdot C_{gNB\text{-}UPF}}{B}$.

*B. MEC@M1*

The propagation delay for the MEC@M1 network deployment can be computed following (4) as:

$$t_{TN,prop\_M1}^X = \frac{d_{gNB\text{-}M1}}{v}, \quad (11)$$

where $d_{gNB\text{-}M1}$ is the distance between the gNB and the M1 node. The transit delay is computed considering that packets are processed and transmitted in the UL TN path by the gNB and M1 node. Then, using (5), it can be computed as:

$$\overline{t_{TN,transit\_M1}^{UL}} = 2 \cdot t_p + \frac{1}{\mu_{gNB}^{UL} - \lambda_{gNB}^{UL}} + \frac{1}{\mu_{M1}^{UL} - \lambda_{M1}^{UL}}. \quad (12)$$

$\lambda_{gNB}^{UL}$ is defined as in (10) and $\mu_{gNB}^{UL} = \frac{\alpha_{gNB\text{-}M1} \cdot C_{gNB\text{-}M1}}{B}$. To calculate $\lambda_{M1}^{UL}$, we consider that each M1 node is connected to a set of $g$ gNBs. Then $\lambda_{M1}^{UL}$ is equal to $\sum_{i=1}^{g} \lambda_{gNB,i}^{UL}$. And following (6) $\mu_{M1}^{UL}$ is equal to $\frac{\alpha_{M1\text{-}UPF} \cdot C_{M1\text{-}UPF}}{B}$. In the DL TN path, the M1 node processes and transmits the packets towards the gNB. The gNB processes the packets and transmits them over the radio interface. Then, the transit delay in the DL is expressed as:

$$\overline{t_{TN,transit\_M1}^{DL}} = 2 \cdot t_p + \frac{1}{\mu_{M1}^{DL} - \lambda_{M1}^{DL}}. \quad (13)$$

In (13), $\mu_{M1}^{DL} = \frac{\alpha_{M1\text{-}gMB} \cdot C_{gNB\text{-}M1}}{B}$ and $\lambda_{M1}^{DL}$ is computed considering that M1 splits traffic in the DL that is addressed to $g$ different gNBs. Then, it can be expressed as $\lambda_{M1}^{DL} = p_k \cdot \lambda_{M1,total}^{DL}$, where the overall traffic rate $\lambda_{M1,total}^{DL}$ that arrives to M1 is equal to the traffic rate departing from the UPF it is connected to (i.e., $\lambda_{UPF}^{DL}$).

*C. MEC@CN*

The propagation delay in the TN for the MEC@CN network deployment can be computed as:

$$t_{TN,prop\_CN}^X = \frac{d_{gNB\text{-}M3}}{v}, \quad (14)$$

where $d_{gNB\text{-}M3}$ is the distance that packets travel from the gNB to the M3 node. In this deployment, packets are processed and transmitted in the UL TN path at a gNB and M1, M2, and M3 nodes. Then, $\overline{t_{TN,transit\_CN}^{UL}}$ can be expressed as:

$$\overline{t_{TN,transit\_CN}^{UL}} = 4 \cdot t_p + \sum_{\forall q \in \{gNB, M1, M2, M3\}} \frac{1}{\mu_q^{UL} - \lambda_q^{UL}}. \quad (15)$$

Traffic aggregation takes place in the UL at the M1, M2 and M3 nodes. In general, we consider that these nodes aggregate traffic from a set of $g$ gNBs, $m_1$ M1 nodes, and $m_2$ M2 nodes, respectively. $\lambda_{gNB}^{UL}$ and $\lambda_{M1}^{UL}$ are computed as shown for MEC@M1, and the arrival rate of traffic to M2 and M3 nodes is computed as $\lambda_{M2}^{UL} = \sum_{i=1}^{m_1} \lambda_{M1,i}^{UL}$ and $\lambda_{M3}^{UL} = \sum_{i=1}^{m_2} \lambda_{M2,i}^{UL}$, respectively.

Besides, the average service rate at M2 and M3 is $\mu_{M2}^{UL} = \frac{a_{M2\text{-}M3} \cdot C_{M2\text{-}M3}}{B}$ and $\mu_{M3}^{UL} = \frac{a_{M3\text{-}UPF} \cdot C_{M3\text{-}UPF}}{B}$, respectively. In the DL TN path, M3, M2, and M1 nodes process and transmit the packets towards the gNB. The gNB then processes them before the packets are transmitted through the radio interface. Then, the transit delay in the DL can be computed as:

$$\overline{t_{TN,transit\_CN}^{DL}} = 4 \cdot t_p + \sum_{\forall q \in \{M1,M2,M3\}} \frac{1}{\mu_q^{DL} - \lambda_q^{DL}}. \quad (16)$$

In this case, the arrival rate of traffic to the M3, M2 and M1 nodes in the DL TN path is computed as $\lambda_{M3}^{DL} = p_k \cdot \lambda_{UPF}^{DL}$ with $k=1, \ldots, m_2$, $\lambda_{M2}^{DL} = p_j \cdot \lambda_{M3}^{DL}$ with $j=1, \ldots, m_1$, and $\lambda_{M1}^{DL} = p_i \cdot \lambda_{M2}^{DL}$ with $i$, $i=1, \ldots, g$. In addition, the average service rate of these nodes is $\mu_{M3}^{DL} = \frac{a_{M3\text{-}M2} \cdot C_{M2\text{-}M3}}{B}$, $\mu_{M2}^{DL} = \frac{a_{M2\text{-}M1} \cdot C_{M1\text{-}M2}}{B}$, and $\mu_{M1}^{DL} = \frac{a_{M1\text{-}gNB} \cdot C_{gNB\text{-}M1}}{B}$, respectively.

### D. Centralized

In the Centralized network deployment, packets pass through the same set of nodes of the TN than in the MEC@CN deployment (i.e., gNB, M1, M2 and M3). Therefore, $t_{TN,prop\_cent}^X$ can be computed as in (14), and $\overline{t_{TN,transit\_cent}^{UL}}$ and $\overline{t_{TN,transit\_cent}^{DL}}$ can be expressed as in (15) and (16), respectively.

## VII. Core Network Latency

The CN latency $l_{CN}$ accounts for the latency that packets experience in the UL ($l_{CN}^{UL}$) and DL ($l_{CN}^{DL}$) CN paths. Like for the TN latency, $l_{CN}^X$ can be expressed as the sum of the propagation delay ($t_{CN,prop}^X$) and the transit delay ($t_{CN,transit}^X$). $t_{CN,prop}^X$ is expressed as in (4) with $d_{CN}$ representing the distance that packets travel through the links that interconnect the CN's nodes, and $v$ the propagation speed over the CN.

This study also estimates $t_{CN,transit}^X$ using a queuing system. Following [11], the nodes of the CN are modeled as an M/D/1 queue. M/D/1 queues are also characterized by $\lambda$ and $\mu$, but in this case $\lambda$ follows a Poisson process and $\mu$ is fixed[3]. This model accounts for the flexibility introduced in 5G with network function virtualization (NFV). It considers NFV nodes placed as gateways between the CN and TN, and between CN and the Internet. These NFV nodes act as UPFs, and they are interconnected through network switches. In the considered network deployments, packets may pass through a (local) UPF of the CN when the V2X AS is hosted by a MEC, or through the UPFs and the intermediate network switches when the V2X AS is hosted in the cloud. When packets pass only through a (local) UPF to reach the MEC, the average transit delay can be estimated as [11]:

$$\overline{t_{CN,transit\_MEC}^X} = \frac{2}{\mu_{UPF}^X} + \frac{\lambda_{UPF}^X}{2(\mu_{UPF}^X)^2 (1-\rho_{UPF}^X)}, \quad (17)$$

where the first term includes the processing ($t_p$) and transmission time delays ($t_{tt}$), and the second term is the queuing delay ($t_q$). When packets need to traverse the CN to be processed in the cloud, the average transit delay is estimated as:

$$\overline{t_{CN,transit\_cloud}^X} = 2 \cdot \left(\overline{t_{CN,transit\_MEC}^X}\right) + \frac{2}{\mu_{UPF}^X} \cdot S. \quad (18)$$

In (18), $S$ is the number of intermediate nodes between the UPFs that act as gateways with the TN and the Internet (i.e., the PSA), and it is computed as:

$$S = \left\lceil \frac{d_{CN}}{d_{CN\_max}} \right\rceil - 1, \quad (19)$$

where $d_{CN\_max}$ is the maximum distance between two nodes of the CN. Note that the first term of (18) considers that packets are only queued at the two UPFs that act as gateways. Intermediate UPFs only add processing and transmission time delays to the CN latency (second term of (18)).

Well-known queuing theory expressions can be used to derive the CN latency distribution function (i.e., $\mathcal{L}_{CN}(t)$) [26]. However, the expressions of $\mathcal{L}_{CN}(t)$ for M/D/1 queuing systems are not tractable [26]. $\mathcal{L}_{CN}(t)$ is then derived in this study using numerical evaluations.

The general expressions for the CN latency are shown above and apply to all 5G network deployments defined. However, the deployments differ in the values of the $t_{CN,prop}^X$ and $t_{CN,transit}^X$ variables that are computed next for each deployment.

#### 1) MEC@gNB

In the MEC@gNB network deployment, packets pass through a local UPF that is physically located close to the gNB and the MEC that hosts the V2X AS. Then, $d_{CN}$ can be approximated to zero, and hence $t_{CN,prop\_gNB}^X = 0$ s. The average transit delay is computed following (17). In this case, since the gNB is connected to the UPF, the arrival rate of traffic to the UPF in the UL (i.e., $\lambda_{UPF}^{UL}$) is equal to the arrival rate of traffic to the gNB it is connected to (i.e., $\lambda_{gNB}^{UL}$, see (10)). The service rate of the UPF in the UL can then be estimated as $\mu_{UPF}^{UL} = \frac{a_{UPF\text{-}AS} \cdot C_{UPF\text{-}AS}}{B}$. In the DL, the UPF forwards the traffic coming from the V2X AS to the gNB it is connected to. Then, $\lambda_{UPF}^{DL}$ is equal to the traffic departure rate from the V2X AS (i.e., $\lambda_{AS}^{DL}$, see details in Section X), and $\mu_{UPF}^{DL} = \frac{a_{UPF\text{-}gNB} \cdot C_{gNB\text{-}UPF}}{B}$.

#### 2) MEC@M1

In the MEC@M1 deployment, packets are also only processed by a local UPF. In this case, the UPF is located at the TN's node M1 where the MEC is also collocated. $d_{CN}$ is then zero, and hence $t_{CN,prop\_M1}^X = 0$ s. The average transit delay is also computed using (17). The difference with MEC@gNB lies in the fact that the UPF is connected to the M1 node. Then, $\lambda_{UPF}^{UL}$ is equal to the arrival rate of traffic to the M1 node that the UPF is connected to (i.e., $\lambda_{M1}^{UL}$, see Section VI.B). In the DL CN path $\mu_{UPF}^{DL}$ is equal to $\frac{a_{UPF\text{-}M1} \cdot C_{M1\text{-}UPF}}{B}$.

---

[3] [11] demonstrates that the traffic departure rate from this queueing system "*is more likely to approach an exponentially distributed random variable than a deterministic value*" when the nodes are not saturated. We consider that this is the case in our analysis, and hence we can estimate that the traffic departure rate is exponentially distributed. We then approximate the traffic arrival rate from the CN to the V2X AS in the UL (see Section X), and from the CN to the TN in the DL (see Section VI) as an exponential distribution.

*3) MEC@CN*

In the MEC@CN deployment, the UPF and MEC are collocated with the TN's node M3. Then, $d_{CN}$ is zero so is the propagation delay. The average transit delay is also estimated using (17). In this case, $\lambda_{UPF}^{UL}$ is equal to the arrival rate of traffic to M3 that the UPF is connected to (i.e., $\lambda_{M3}^{UL}$, see Section VI.C). In the DL CN path, $\mu_{UPF}^{DL}$ is $\frac{a_{UPF\text{-}M3} \cdot C_{M3\text{-}UPF}}{B}$.

*4) Centralized*

In the Centralized network deployment, packets are routed through the CN to be processed in the cloud. $t_{CN,prop\_cent}^{X}$ can then be estimated considering that $d_{CN}$ is the distance from the M3 node to the PSA UPF that serves as gateway to the Internet. The CN average transit delay is computed using (18). Note that in this case $\lambda_{UPF}^{UL} = \lambda_{M3}^{UL}$ and $\mu_{UPF}^{UL} = \frac{a_{UPF\text{-}S1} \cdot C_{UPF\text{-}S1}}{B}$ because the arrival rate of traffic to the UPF that acts as gateway between the TN and CN is equal to the arrival rate of traffic to the M3 node, and this UPF forwards the traffic towards the first intermediate UPF (named *S1*). In the DL, $\lambda_{UPF}^{DL} = \lambda_{AS}^{DL}$ and $\mu_{UPF}^{DL} = \frac{a_{UPF\text{-}Ss} \cdot C_{Ss\text{-}UPF}}{B}$ because the arrival rate of traffic to the PSA UPF is equal to the traffic departure rate from the V2X AS, and this UPF forwards the traffic in the DL towards the $s^{th}$ intermediate UPF node (named *Ss*).

## VIII. INTERNET LATENCY

This study also estimates the latency between the CN (i.e., the PSA UPF) and the V2X AS when the V2X AS is located at the cloud. We refer to this latency as the Internet latency and note it as $l_{UPF\text{-}AS}$. Note that $l_{UPF\text{-}AS}$ is only present in the Centralized network deployment. Like in the previous analysis, $l_{UPF\text{-}AS} = l_{UPF\text{-}AS}^{UL} + l_{UPF\text{-}AS}^{DL}$. This work models the Internet latency using the empirical study reported in [27] that characterizes the CDF (i.e., $\mathcal{L}_{UPF\text{-}AS\_cent}^{X}(t)$) of the round-trip time observed between source-target Internet nodes. The model accounts for the transmission delay, propagation delay, and queuing/processing time (at intermediate routers/switches) between a pair of nodes in the Internet path. The Internet latency measurements reported in [27] correspond to the minimum round-trip time value observed between any pair of source-target Internet nodes. We consider from [27] the best-case (i.e., lower latency) scenario where these nodes are located in the same country (Fig. 4). This is also the most likely scenario for a V2N2V connection involving two neighboring vehicles. This scenario represents the case in which the vehicles that are communicating and the V2X AS that processes their traffic are in the same country. V2N connections consider half of the round-trip time.

## IX. PEERING-POINTS LATENCY

Different MNOs are usually connected via remote (or public) peering points at Internet exchange points. MNOs can also connect directly with each other via local (or private) peering points. We model the latency introduced by these peering points using the measurements in [28] that characterize the CDF of the latency for the local (i.e., $\mathcal{L}_{pp\text{-}local}^{X}(t)$) and remote peering points (i.e., $\mathcal{L}_{pp\text{-}remote}^{X}(t)$) (see Fig. 5).

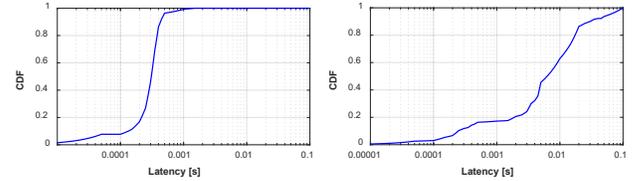

Fig. 5. Latency distribution of local (left) and remote (right) peering points.

## X. V2X AS LATENCY

A V2X AS may add additional latency ($l_{AS}$) due to, for example, the required data processing at the AS for a V2X service and/or the forwarding of the received packets. We consider in this study that the V2X AS acts as a forwarder as in [10]. This results in that the traffic departure rate from the V2X AS can be expressed as $\lambda_{AS}^{DL} = M \cdot \lambda_{AS}^{UL}$, where $M$ is the number of copies of the packet received at the V2X AS that are forwarded in the DL (e.g., $M$ is equal to 1 when DL packets are broadcasted, and higher than 1 when forwarded using unicast transmissions) and $\lambda_{AS}^{UL}$ can be approximated to $\lambda_{UPF}^{UL}$ (see Section VII). In [10], the authors derive the following expression to compute the V2X AS latency at the V2X AS when it acts as a forwarder:

$$l_{AS} = \eta \cdot B \cdot \theta / F, \quad (20)$$

where $F$ is the processing capacity of the V2X AS (in cycles/s), $B$ is the packet size (in bits), $\theta$ is the cycles/bit necessary for the forwarding of the received V2X packets, and $\eta$ is the number of processed packets. As in [10], $\theta$ is modeled using a continuous uniform distribution $U(100, 300)$[4]. Then, $\overline{l_{AS}}$ and $\mathcal{L}_{AS}(t)$ are computed as in (20) using the mean value and CDF that characterizes $\theta$, respectively. $\eta$ depends on the time window established to process packets at the AS. We consider that the time window is set to the radio transmission time or slot time duration ($t_{tt}$), and its value depends on the 5G NR numerology. Then, the average number of packets received at the V2X AS in $t_{tt}$ seconds can be approximated as $\eta_{tt} = \lambda_{AS}^{UL} \cdot t_{tt}$.

## XI. EVALUATION SCENARIO

The 5G E2E latency model presented in this paper is utilized to evaluate the capabilities of 5G to support V2X services via a V2N2V connection. Note that the evaluation of a V2N connection could also be conducted by taking the UL or DL components of the 5G E2E latency model. Following the studies reported by the ITU-T Study Group in [18][29], the evaluation considers a hierarchical optical transport network architecture. The transport network is made of 3 levels of multiplexing nodes (i.e., M1, M2 and M3) that connect a number of gNBs to the CN. Following [25], we consider that each M1 node multiplexes traffic from 6 different gNBs. A group of 6 of these M1 nodes connected in a ring topology form an access ring, and each access ring serves a total of 36 gNBs. M1 nodes are point-to-point connected to the M2 nodes. Each M2 node serves a total of 4 access rings and 144 gNBs. 6 M2 nodes are connected in a ring topology to form an aggregation ring. The M2 nodes are also point-to-point connected to M3 nodes. Each M3 node serves a total of 2 aggregation rings and 1728 gNBs. Following a reference deployment from the ITU-T Study Group [29], the distance between the gNB and the M1 nodes is set to 3 km, between M1 and M2 nodes to 12 km, and between M2 and M3 nodes to 60 km. [25] identifies link capacities of 10 Gb/s for the gNB-M1 links, 300 Gb/s for the M1-M2 links, and 6 Tb/s for the M2-M3 links. Following reports from the ITU-T Study Group [30][31], we consider a CN distance of 200 km. We set the link capacity between the CN nodes to 6 Tb/s following the study reported in [32].

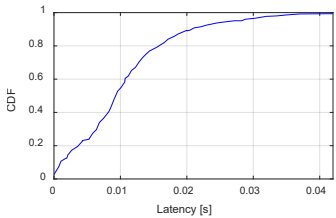

Fig. 4. Minimum round-trip time Internet latency (source-target pairs in Italy) [27]. Values for other countries or the whole Europe can also be found in [27].

---

[4] $\theta$ is on average 200 cycles/bit. Since the arrival rate of traffic to the TN and CN follows an exponential distribution, we approximate the AS processing delay with an exponential distribution characterized by the average value of $\theta$.

TABLE III. V2X SERVICE REQUIREMENTS

|  | Max. E2E Latency | Reliability |
|---|---|---|
| Low LoA (LLoA) | 25 ms | 90 % |
| High LoA (HLoA) | 10 ms | 99.99 % |

Our study is based on the V2X scenarios identified by the 3GPP in [33]. The 3GPP defines V2X services with low or high level of automation (LLoA or HLLoA) (see Table III). LLoA services require that 90% of the transmitted packets are successfully delivered in less than 25 ms. HLoA services require that 99.99% of the transmitted packets are received in less than 10 ms. This study does not focus on specific V2X services but considers instead a rate $\lambda_{gNB}^{UL}$ of V2X packets arriving at each gNB in the scenario. $\lambda_{gNB}^{UL}$ is set equal to {1040, 2080, 4160, 5200, 6240, 8320, 10400, 20800, 31200, 41600} packets/sec to emulate different traffic densities (i.e., {10, 20, 40, 60, 80} vehicles/km) and loads (i.e., $T_p$ = {10, 50} packets/s). The set of values of $\lambda_{gNB}^{UL}$ has been derived following 3GPP guidelines for the evaluation of advanced V2X services that consider a highway scenario with 3 lanes per direction and a 5G NR cell radius of 0.866 Km. Without loss of generality, we consider the same average $\lambda_{gNB}^{UL}$ per gNB.

## XII. LATENCY ANALYSIS

This section evaluates the radio network, TN, CN, Internet, peering points and V2X AS latencies using a reference scenario (unless otherwise specified) with $\lambda_{gNB}^{UL}$=2080 pkts/s. Results are compared, when available, with measurements reported in the literature. We should note that the challenge and cost associated with building a complete and accurate E2E network simulator or deploying different 5G networks for field experiments limits the availability of reference measurements or results.

### A. Radio network

We use the radio latency results obtained in the simulations and model we derived in [16]. Fig. 6 shows the average V2N2V (i.e., UL+DL) radio latency ($l_{radio}$) with a solid lane, as well as its 90th (left) and 99.99th (right) percentiles with a dashed line, as a function of the traffic density. Fig. 6 reports $l_{radio}$ for a scenario where $T_p$ = {100, 20} ms, and for a common reference 5G NR configuration with a SCS = 30 KHz, 14 OFDM symbols, and a cell bandwidth of 20 MHz for both UL and DL [34]. Table IV shows the UL + DL radio latency values for the selected configuration. The results for the LLoA and HLoA V2X services correspond to the 90th and 99.99th percentile of the measured $l_{radio}$, and they are obtained using a transmission mode that guarantees a reliability of 90% (i.e., MCS Table 1 in [21]) and $10^{-5}$ (i.e., MCS Table 3 in [21]). Results are reported for different values of $\lambda_{gNB}^{UL}$. In Table IV '-' represents a combination that cannot satisfy the latency requirements of the selected service[5].

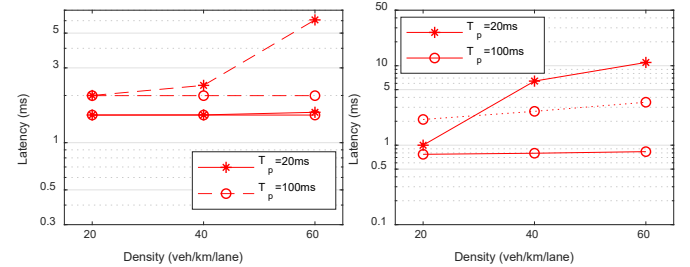

Fig. 6. Average (solid line), and 90th percentile (left-dashed line) and 99.99th percentile (right-dashed line), of the V2N2V (UL + DL) radio latency.

TABLE IV. UL+DL RADIO LATENCY IN MS [16]

| LoA | $\lambda_{gNB}^{UL}$ /pkts/s/ | | | | | | | | | |
|---|---|---|---|---|---|---|---|---|---|---|
|  | 1040 | 2080 | 4160 | 5200 | 6240 | 8320 | 10400 | 20800 | 31200 | 41600 |
| Low | 2.00 | 2.00 | 2.00 | 2.00 | 2.00 | 2.00 | 2.00 | 2.32 | 6.07 | - |
| High | 2.60 | 2.77 | 3.08 | 3.58 | 3.58 | 4.55 | - | - | - | - |

### B. Transport network

The transport network latency includes propagation delay $t_{TN\_prop}$ and transit delay $t_{TN\_transit}$. Following (4), $t_{TN\_prop}$ depends on the total distance $d_{TN}$ that packets travel through the TN, and the propagation speed $v$ over the TN. For the considered optical TN, $v$ can be approximated to 200000 km/s, and $d_{TN}$ is {0, 3, 75, 75} km for the {MEC@gNB, MEC@M1, MEC@CN, Centralized} deployments. Table V computes the resulting $t_{TN,prop}^{UL} + t_{TN,prop}^{DL}$ for the different deployments. $t_{TN\_prop}$ increases with the distance that V2X packets need to travel to reach the V2X AS.

TABLE V. UL + DL PROPAGATION AND PROCESSING DELAYS IN MS

| Metric | Deployment | | | |
|---|---|---|---|---|
|  | MEC@gNB | MEC@M1 | MEC@CN | Centralized |
| $t_{TN,prop}$ | - | 0.03 | 0.75 | 0.75 |
| $t_p$ | 0.4 | 0.8 | 1.6 | 1.6 |

---

[5] To compute the radio latency $l_{radio}$, we consider that a transmitter/vehicle drops a packet if it has not been transmitted when a new one is generated [16]. Table IV reports '-' when the percentage of dropped packets is higher than 10% and 0.01% for the LLoA and HLoA V2X services since the 90th percentile and 99.99th percentile of $l_{radio}$ cannot be computed (Fig. 6).

The $t_{TN,transit}$ depends on the processing $t_p$ and queuing delays $t_q$ at the TN's nodes. This study approximates the TN's nodes $t_p$ to 0.2 ms following [35]. $t_p$ increases then with the number of TN nodes that process the packets (Table V), which depends on the specific network deployment. For example, in the MEC@gNB network deployment, $t_p$ = 0.4 ms since the gNB processes the packets in the UL and DL. In the MEC@CN and Centralized network deployments, $t_p$ = 1.6 ms since the packets pass through the gNB, M1, M2 and M3 nodes both in the UL and DL. The analysis of $t_q$ does not only depend on the deployment, but also on the V2X traffic load that traverses the nodes of the TN, and the link capacities allocated to support the traffic of the V2X service. For example, Fig. 7 shows the distribution of the UL queuing delay for the MEC@M1 network deployment when the arrival rate of traffic $\lambda_{gNB}^{UL}$ is 2080 pkts/s[6]. Results are reported in Fig. 7 for different values of $\alpha$ or ratio of the links capacities that is allocated to support the traffic of the V2X service in UL and DL. The results in Fig. 7 show that $t_q$ decreases with the increasing values of $\alpha$. For example, the 90th percentile of $t_q$ is 1.1914 ms when $\alpha$ is set to 0.001. The 90th percentile of $t_q$ decreases to 0.0602 ms and 0.0574 ms when $\alpha$ increases to 0.01 and 0.1, respectively. Higher values of $\alpha$ directly contribute to reducing $t_q$ and hence the TN latency. However, this is at the cost of reserving a larger portion of the link capacities to support the specific V2X services. These results show the existing tradeoff between $t_q$ and $\alpha$, and the need to properly set $\alpha$. Identifying the value of $\alpha$ calls for a proper dimensioning that considers both the V2X traffic that the network deployment has to support and the V2X service requirements.

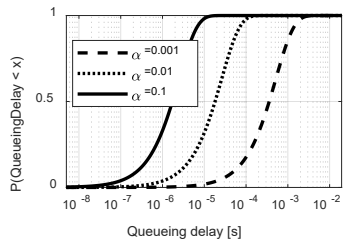

Fig. 7. UL queuing delay $t_q$ as a function of $\alpha$ (MEC@M1)

Table VI shows the sum of $t_{TN,prop}$, $t_p$ and $t_q$ for the considered 5G network deployments. Table VI shows the average, 90th and 99.99th percentiles of the TN latency. The reported results show how the TN latency decreases as the 5G network deployment uses a V2X AS closer to the radio interface. For example, when $\alpha$=0.1, $\overline{l_{TN}}$ is 2.355 ms for the deployments MEC@CN and Centralized in which the packets pass through all nodes of the TN (i.e., gNB, M1, M2 and M3), and $\overline{l_{TN}}$ decreases to 0.835 ms in the MEC@M1 deployment in which packets only pass through the gNB and M1 TN's nodes. For the MEC@gNB deployment, $\overline{l_{TN}}$ reduces to 0.402 ms. Similar trends are reported for the 90th and 99.99th percentile of the TN latency. Table VI also shows the impact of $\alpha$ on the TN latency. The $\alpha$ and $t_q$ trade-offs highlighted in Fig. 7 result in this case in that the MEC@CN and Centralized network deployments cannot support the V2X traffic when $\alpha$ is set to 0.001; in this case, the utilization $\rho$ is higher than 1 (Section VI). Note also that the 99.99th percentile of the TN latency for MEC@M1 is ~10.3 ms, while HLoA requires that 99.99% of the packets experience an E2E latency below 10 ms. The 99.99th percentile of the TN latency decreases to 1.3 ms and 0.875 ms when $\alpha$ is set to 0.01 and 0.1, respectively. Note that the results reported in Table VI are in line with the TN latency measurements reported in [23] for a 5G network in London.

TABLE VI. TRANSPORT NETWORK LATENCY IN MS

| Deployment | Metric | $\alpha$ =0.001 | $\alpha$ =0.01 | $\alpha$ =0.1 |
|---|---|---|---|---|
| MEC@gNB | $\overline{l_{TN}}$ | 0.908 | 0.425 | 0.402 |
|  | 90th percentile ($l_{TN}$) | 1.571 | 0.458 | 0.410 |
|  | 99.99th percentile ($l_{TN}$) | 5.083 | 0.633 | 0.422 |
| MEC@M1 | $\overline{l_{TN}}$ | 1.856 | 0.881 | 0.835 |
|  | 90th percentile ($l_{TN}$) | 3.192 | 0.949 | 0.841 |
|  | 99.99th percentile ($l_{TN}$) | 10.279 | 1.304 | 0.875 |
| MEC@CN Centralized | $\overline{l_{TN}}$ | * | 2.402 | 2.355 |
|  | 90th percentile ($l_{TN}$) | * | 2.471 | 2.361 |
|  | 99.99th percentile ($l_{TN}$) | * | 2.833 | 2.396 |

*This value of $\alpha$ is not sufficient to support the V2X traffic.

*C. Core network*

The CN latency includes propagation delay $t_{CN,prop}$, processing delay $t_p$ and queuing delay $t_q$. The only 5G network deployment in which packets travel through the CN is the Centralized one. Considering that the CN distance is 200 km and an optical CN ($v$ = 200000 km/s), $t_{CN,prop}$ is 2 ms for the Centralized deployment. Table VII reports the CN latency for all deployments. As it was also highlighted in [36] through measurements in Helsinki, the obtained results show that the CN latency is mostly dominated by the $t_{CN,prop}$. This is the case because $t_p$ and $t_q$ are inversely related with the capacity of the CN links (see (17) and (18)) which are of the order of Tb/s for the considered 5G CN; in the TN, the link capacities are of the order of Gb/s. Table VII also highlights the latency and $\alpha$ tradeoffs previously discussed. Table VII also shows that the Centralized deployment cannot satisfy the HLoA and LLoA V2X service requirements when $\alpha$= 0.001.

TABLE VII. CORE NETWORK LATENCY IN MS

| Deployment | Metric | $\alpha$ =0.001 | $\alpha$ =0.01 | $\alpha$ =0.1 |
|---|---|---|---|---|
| MEC@gNB MEC@M1 MEC@CN | $\overline{l_{CN}}$ | 0.0016 | 0.0001 | 0.00001 |
|  | 90th percentile ($l_{CN}$) | 0.0008 | 0.00008 | 0.00001 |
|  | 99.99th percentile ($l_{CN}$) | 0.0008 | 0.00008 | 0.00001 |
| Centralized | $\overline{l_{CN}}$ | * | 2.0006 | 2.0000 |
|  | 90th percentile ($l_{CN}$) | * | 2.0006 | 2.0001 |
|  | 99.99th percentile ($l_{CN}$) | * | 2.0007 | 2.0001 |

*This value of $\alpha$ is not sufficient to support the V2X traffic.

*D. Internet*

The latency introduced by Internet connections only intervenes in the Centralized network deployment. Based on [27] and Fig. 4, the 90th and 99.99th percentiles of the Internet latency are equal to 21 ms and 43 ms, respectively. This shows that the contribution of the Internet latency makes the Centralized deployment not suitable to satisfy with V2N2V connections the requirements of the HLoA V2X service (Table III). The 90th percentile of the Internet latency does not exceed the latency requirement of the LLoA V2X service. However, the sum of all other latency contributions of V2N2V E2E connections must not surpass 4 ms in Centralized deployments to satisfy the latency requirement of this service.

---

[6] Similar trends are observed for the other deployments and loads.

### E. Peering points

The latency introduced by peering points connections between MNOs is modeled using the empirical measurements reported in [28] and shown in Fig. 5. Table VIII shows the average, 90th and 99.99th percentiles of the latency for the remote and local peering points. The 90th and 99.99th percentiles of the remote peering-points latency are above the latency requirements for the LLoA and HLoA V2X services (Table III). The results highlight the challenge to support these V2X services using V2N2V connections in multi-MNO scenarios using remote peering points. Table VIII shows that the use of local peering-points significantly reduces the latency and improves the prospects to support V2X services using 5G-based V2N2V connections in multi-MNO scenarios.

TABLE VIII. PEERING POINTS LATENCY IN MS

|        | $\overline{l_{pp}}$ | 90th percentile ($l_{pp}$) | 99.99th percentile ($l_{pp}$) |
|--------|------|-------|-------|
| Remote | 13.001 | 29.867 | 99.212 |
| Local  | 0.306 | 0.431 | 1.493 |

### F. V2X AS

The contribution of the V2X AS latency to the 5G E2E latency is computed using (20). We use as a reference the features of processors that are typically used in cloud and MEC platforms. We consider for the deployment of the AS on the cloud processors that includes 28 cores (56 threads) and a maximum core frequency of 4.30GHz. For the MEC deployments, processors with 24 cores (48 threads) and a maximum core frequency of 3.6GHz are considered [37]. Following [38], the number of processors available to each application located at the cloud range between 23 and 110. For the MEC deployments at CN and M1, we consider that there are 4 processors available for each application [39]. For the MEC deployment at the gNB there are 2 processors available for each application [40]. Considering these deployments, Table IX reports the V2X AS latency for the different 5G network deployments when $\lambda_{gNB}^{UL}$ is 2080 and 41600 pkts/s. The results show that the V2X AS latency increases with $\lambda_{gNB}^{UL}$.

To avoid that packets backlog in the AS's queue, the V2X AS needs to process the received packets in less than the radio transmission time $t_{tt}$ (i.e., $\overline{l_{AS}} \leq t_{tt}$). If $t_{tt}$=0.5 ms (i.e., SCS = 30 kHz), packets backlog in the MEC@CN and Centralized deployments when $\lambda_{gNB}^{UL}$ > 2080 pkts/s. In this case, a re-dimensioning of the processing power of the AS needs to be conducted, and this is analyzed in the following section.

TABLE IX. V2X AS LATENCY IN MS*

|  | MEC@gNB | | MEC@M1 | | MEC@CN | | Centralized | |
|---|---|---|---|---|---|---|---|---|
| $\lambda_{gNB}^{UL}$ [pkts/s] | 2080 | 41600 | 2080 | 41600 | 2080 | 41600 | 2080 | 41600 |
| $\overline{l_{AS}}$ | 0.0027 | 0.0031 | 0.0048 | 0.0917 | 1.320 | 2.640 | 0.035 – 0.165 | 0.689 – 3.295 |
| 90th perc.($l_{AS}$) | 0.039 | 0.042 | 0.068 | 0.128 | 1.843 | 3.662 | 0.0482 – 0.231 | 0.701 – 4.633 |
| 99.99th perc. ($l_{AS}$) | 0.042 | 0.046 | 0.073 | 0.137 | 1.980 | 3.955 | 0.0517 – 0.247 | 0.750 – 4.940 |

*MEC@{gNB, M1, CN} and Centralized uses {2, 4, 4, 110-23} processors.

## XIII. DIMENSIONING

The conducted analysis has shown the need to adjust the V2X AS processing power and the ratio $\alpha$ of the link capacities allocated to support the traffic of V2X services and meet their latency requirements. This section focuses on the dimensioning of these two factors.

### A. V2X AS processing power

The processing power of the V2X AS needs to be adjusted so that packets are not backlogged at the AS queue. To this aim, it is necessary that the $\eta_{tt} = \lambda_{AS}^{UL} \cdot t_{tt}$ packets that arrived at the V2X AS are processed in less than $t_{tt}$ seconds, which is equivalent to $\overline{l_{AS}} \leq t_{tt}$. Table X reports the minimum number of processors necessary to avoid packets backlogged, which are in line with the deployments reported in [38]-[40]. These numbers are obtained using (20). Table X shows that the network deployment, and hence the location of the AS, has a high impact on the number of processors needed, and considerably increases when the AS is installed at the CN or at the cloud. We should note that although the MEC@CN and Centralized deployments require more processors per AS node, they also need less AS nodes to serve a geographical area.

TABLE X. MINIMUM NUMBER OF PROCESSORS NECESSARY TO AVOID PACKETS BACKLOG AT THE V2X AS

| $\lambda_{gNB}^{UL}$ | Deployments | | | |
|---|---|---|---|---|
|  | MEC@gNB | MEC@M1 | MEC@CN | Centralized |
| 2080 | 1 | 1 | 6 | 4 |
| 41600 | 1 | 1 | 212 | 152 |

### B. Dimensioning of α

The capacity of the transport and core networks to satisfy the requirements of V2X services depend on the fraction of their link capacities $\alpha$ allocated for serving these services. Augmenting $\alpha$ reduces the TN and CN latency but reduces the capacity to serve other users/services. This section identifies the minimum $\alpha$ ($\alpha_{min}$) necessary to satisfactorily support the V2X services using the framework in (21)-(26). Without loss of generality, we focus on V2N2V connections and consider the same $\alpha$ is utilized for all the links of the TN and CN both in UL and DL (see (22)). Equation (23) establishes that the utilization $\rho = \lambda/\mu$ of the TN and CN nodes must be lower than 1 so that packets are not backlogged at their queues. In (23), $N$ refers to the set of TN and CN nodes that participate in the forwarding of packets in the UL and DL; $N$ depends on the specific network deployment. Then, (24) indicates that the E2E latency $l_{E2E}$ must satisfy the latency requirements $L_{REQ}$ of the considered V2X services (Table III). (24) establishes that the 90th and 99.99th percentiles of $l_{E2E}$ need to be lower than or to 25 ms and 10 ms for the LLoA and HLoA V2X services, respectively. Finally, (25) and (26) limit the possible values of $\alpha$.

$$\text{Minimize} \quad \alpha_{UL} \tag{21}$$
$$\text{Subject to:} \quad \alpha_{UL} = \alpha_{DL} \tag{22}$$
$$\lambda_i^X < \mu_i^X = \frac{\alpha_i \cdot C_{i\text{-}i+1}}{B}, \ i \in N \tag{23}$$
$$l_{E2E} \leq L_{REQ} \tag{24}$$
$$\alpha_{UL} + \alpha_{DL} < 1 \tag{25}$$
$$0 < \alpha_{UL} < 0.5 \tag{26}$$

Fig. 8.a depicts $\alpha_{min}$ as a function of $\lambda_{gNB}^{UL}$. $\alpha_{min}$ is derived using (21)-(26) and the 5G E2E latency $l_{E2E}$ is computed using the models in Sections IV-X. To estimate $l_{E2E}$, we use the radio latency $l_{radio}$ results reported in Table IV (Section XII.A). We also consider that the V2X AS latency $\overline{l_{AS}}$ is upper bounded by $t_{tt}$ (i.e., 0.5 ms since SCS = 30 kHz). The 90$^{th}$ and 99.99$^{th}$ percentiles of $l_{AS}$ are then 0.6977 ms and 0.7499 ms, respectively (see Section X). Fig. 8.a considers the MEC@M1 network deployment; similar trends as those visible in Fig. 8.a have been observed for other deployments and configurations. Fig. 8.a shows that higher values of $\alpha_{min}$ are necessary as the network traffic load $\lambda_{gNB}^{UL}$ increases. Higher values of $\alpha_{min}$ are also necessary with higher V2X service requirements (HLoA vs. LLoA). Fig. 8.a shows that it is not possible to derive an $\alpha_{min}$ that satisfies the constraints in (21)-(26) for the HLoA V2X service when $\lambda_{gNB}^{UL} > 8320$ pkts/s. This is the case because $l_{radio}$ does not satisfy the HLoA latency requirements (represented with the symbols '-' in Table IV, Section XII.A). Similarly, $\alpha_{min}$ cannot be obtained when $\lambda_{gNB}^{UL} > 31200$ pkts/s for the LLoA V2X service.

Fig. 8.b depicts $\alpha_{min}$ as a function of the 5G network deployment when $\lambda_{gNB}^{UL} = 2080$ pkts/s. Results are reported for single and multiple MNO scenarios. We should note that multi-MNO scenarios add the local peering point latency $l_{pp}$ to $l_{E2E}$ (Table VIII). Fig. 8.b shows that both LLoA and HLoA V2X services require higher values of $\alpha_{min}$ as the V2X AS is located closer to the CN. This is the case because as the V2X AS is located closer to the CN, TN nodes need to multiplex the V2X traffic from a higher number of gNBs, which requires higher link capacities. Fig. 8.b shows that satisfying the latency requirements of the LLoA V2X service requires the same $\alpha_{min}$ in single or multi-MNO scenarios. This is the case because $\alpha_{min}$ is restricted in these scenarios by the TN and CN's nodes utilization expressed in (23), and $l_{E2E}$ is lower than the $L_{REQ}$ of the LLoA V2X service (i.e., 25 ms) for the derived $\alpha_{min}$. On the other hand, Fig. 8.b shows that different values of $\alpha_{min}$ are necessary under the single- and multi-MNO scenarios to satisfy the latency requirements of the HLoA V2X service. In this case, $\alpha_{min}$ values are restricted by the limit imposed on the $l_{E2E}$ by a stricter latency requirement of $L_{REQ}$=10 ms (expressed in (24)) both in single- and multi-MNO scenarios. Therefore, a higher $\alpha_{min}$ is needed to compensate for the $l_{pp}$ that is added to $l_{E2E}$ in the multi-MNO scenario.

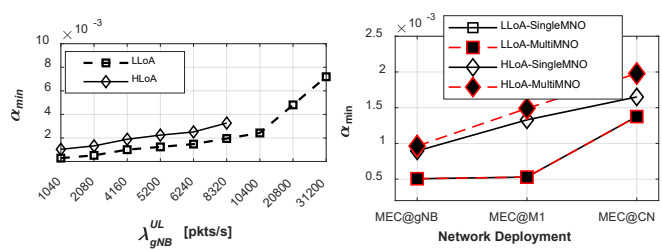

Fig. 8. Derived $\alpha_{min}$ for a) MEC@M1 (left) and b) $\lambda_{gNB}^{UL}$ =2080 (right).

## XIV. END-TO-END LATENCY ANALYSIS

This section analyzes the E2E latency of V2N2V for the 5G network deployments under evaluation. We then analyze the capacity of each deployment to satisfy the latency requirements of the LLoA and HLoA V2X services considered in this study (Table III). This section reports results obtained with the radio latency values reported in Table IV. The evaluation in this section focuses on the values of $\lambda_{gNB}^{UL}$ resulting in a radio latency below the requirements of the selected service, which happens for light to moderate traffic loads (i.e., $\lambda_{gNB}^{UL}$<8320 pkts/s for HLoA and $\lambda_{gNB}^{UL}$<31200 pkts/s for LLoA). In multi-MNO scenarios, we consider that MEC-based network deployments use local peering points, while the Centralized one rely on remote peering points. We evaluate the E2E latency considering the existing tradeoff on the allocated $\alpha$ (Section XIII.B). Then, we have conducted and exhaustive dimensioning analysis following the procedure presented in Section XIII.B for all 5G network deployments and configurations to set $\alpha$ = {0.001, 0.01}. This analysis showed that 20% of the 5G network deployments and configurations analyzed under this study can support the E2E latency requirements of the selected V2X services when $\alpha$=0.001. This value increases to 95% when $\alpha$=0.01. Results in this section are also compared, when available, with measurements reported in the literature. It is though important to note that details on the considered network deployments are not commonly reported.

### A. MEC@gNB

Table XI reports the values of $l_{radio}$, $l_{TN}$, $l_{CN}$, $l_{AS}$, and $l_{pp}$, as well as E2E latency in single-MNO ($l_{E2E}^{singleMNO}$) and multi-MNO ($l_{E2E}^{multiMNO}$) scenarios for the MEC@gNB network deployment. Cells colored in green and red identify the configurations of $\alpha$ and $\lambda_{gNB}^{UL}$ that result in E2E latency values below and above, respectively, the requirements of the corresponding service (Table III). The mark '-' is used when the traffic of the V2X service cannot be supported under this configuration of $\alpha$ and $\lambda_{gNB}^{UL}$. Table XI shows that a deployment with $\alpha$ = 0.001 and $\lambda_{gNB}^{UL}$ = 2080 pkts/s can support the latency requirements of the LLoA V2X service under single-MNO scenarios since $l_{E2E}^{singleMNO}$ is 4.265 ms. In multi-MNO scenarios, the latency of the local peering point (i.e., $l_{pp} = 0.431$ ms) is added and the E2E latency increases to $l_{E2E}^{multiMNO}$ = 4.696 ms, which is still below the requirement of 25 ms for this service.

Table XI shows that when $\lambda_{gNB}^{UL}$ increases to 8320 pkts/s, the TN cannot process all the V2X traffic in less than $L_{REQ}$ if $\alpha$ = 0.001; this is also the case when $\lambda_{gNB}^{UL}$=31200 pkts/s and then it is not included in Table XI for $\alpha$ = 0.001. An analysis conducted using the optimization framework in Section XIII.B shows that the minimum value of $\alpha_{min}$ necessary in this scenario to satisfy $L_{REQ}$ is 0.0019. Using this $\alpha_{min}$, $l_{E2E}^{singleMNO}$ and $l_{E2E}^{multiMNO}$ are equal to 20.49 ms and 20.92 ms, respectively, and hence satisfy $L_{REQ}$ for the LLoA V2X service. Table XI shows that assigning an $\alpha$ significantly higher than $\alpha_{min}$ (i.e., $\alpha$ = 0.01) results in E2E latency values lower than $L_{REQ}$ even for the highest traffic loads. For example, $l_{E2E}^{singleMNO}$ and $l_{E2E}^{multiMNO}$ are equal to 7.435 ms and 7.866 ms, respectively, when $\lambda_{gNB}^{UL}$ = 31200 pkts/s and $\alpha$ = 0.01.

Table XI shows that supporting the HLoA V2X service under MEC@gNB is more challenging when $\alpha$ = 0.001. In this case, only the single-MNO scenario with the lowest traffic load (i.e., $\lambda_{gNB}^{UL}$ = 2080 pkts/s) is supported. Fulfilling the HLoA latency requirements in the multi-MNO scenario could be achieved using an $\alpha_{min}$ = 0.00102 that reduces $l_{TN}$ and $l_{CN}$. When $\lambda_{gNB}^{UL}$ increases to 8320 pkts/s, the necessary $\alpha_{min}$ to satisfy the latency requirements of the HLoA V2X service increases to 0.0026 and 0.0029 for the single-MNO and multi-MNO scenarios, respectively.

The E2E latency values shown in Table XI are well aligned with the results reported in [7] for light to moderate traffic loads. In particular, [7] reports E2E latency values that range from ~3 ms to ~20 ms in a 5G standalone network deployment where the MEC is colocated with the gNB and the radio interface is also configured with 20 MHz bandwith (the considered SCS is 60 kHz). Details about the allocated $\alpha$ or the processing capabilities of the AS are not reported in [7].

TABLE XI. E2E LATENCY (IN MS) FOR THE MEC@GNB DEPLOYMENT.

|  | $\alpha = 0.001$ |  |  |  | $\alpha = 0.01$ |  |  |  |  |
|---|---|---|---|---|---|---|---|---|---|
|  | LLoA |  | HLoA |  | LLoA |  |  | HLoA |  |
| $\lambda_{gNB}^{UL}$ | 2080 | 8320 | 2080 | 8320 | 2080 | 8320 | 31200 | 2080 | 8320 |
| $l_{radio}$ | 2.00 | 2.00 | 2.77 | 4.55 | 2.00 | 2.00 | 6.07 | 2.77 | 4.55 |
| $l_{TN}$ | 1.571 | - | 5.083 | - | 0.458 | 0.470 | 0.665 | 0.633 | 0.680 |
| $l_{CN}$ | 0.001 | 0.001 | 0.001 | 0.014 | 0.0001 | 0.0001 | 0.001 | 0.0001 | 0.001 |
| $l_{AS}$ | 0.6977 |  | 0.7499 |  | 0.6977 |  |  | 0.7499 |  |
| $l_{E2E}^{singleMNO}$ | 4.265 | - | 8.603 | - | 3.152 | 3.164 | 7.435 | 4.153 | 5.979 |
| $l_{pp}$ | 0.431 |  | 1.493 |  | 0.431 |  |  | 1.493 |  |
| $l_{E2E}^{multiMNO}$ | 4.696 | - | 10.096 | - | 3.582 | 3.595 | 7.866 | 5.646 | 7.473 |

### B. MEC@M1

Table XII shows similar trends for the E2E latency in the MEC@M1 network deployment as those explained for the MEC@gNB deployment. However, Table XII shows higher latency values for the $l_{TN}$ and $l_{CN}$ in MEC@M1 than in MEC@gNB. This is due to the larger distances that packets need to travel through TN and CN and the higher number of packets that are processed by the TN and CN's nodes as the V2X AS moves closer to the CN. This results in higher values of $l_{E2E}^{singleMNO}$ and $l_{E2E}^{multiMNO}$ reported in Table XII compared to Table XI It is then necessary to increase $\alpha$ to satisfy the latency requirements in certain scenarios. For example, unlike the MEC@gNB deployment, the HLoA latency requirements cannot be satisfied in MEC@M1 when $\alpha = 0.001$ even under the lowest traffic load (i.e., $\lambda_{gNB}^{UL}$ = 2080 pkts/s). Satisfying such requirements requires increasing $\alpha$ to a minimum value of 0.0014 and 0.0015 for the single-MNO and multi-MNO scenarios, respectively. These values increase to 0.0033 and 0.004, respectively, when the traffic load increases to $\lambda_{gNB}^{UL}$ = 8320 pkts/s. These results basically show that as the V2X AS is located closer to the CN it is necessary to reserve more of the link capacity (i.e., increase $\alpha$) to satisfy the V2X latency requirements.

TABLE XII. E2E LATENCY (IN MS) FOR THE MEC@M1 DEPLOYMENT.

|  | $\alpha = 0.001$ |  |  |  | $\alpha = 0.01$ |  |  |  |  |
|---|---|---|---|---|---|---|---|---|---|
|  | LLoA |  | HLoA |  | LLoA |  |  | HLoA |  |
| $\lambda_{gNB}^{UL}$ | 2080 | 8320 | 2080 | 8320 | 2080 | 8320 | 31200 | 2080 | 8320 |
| $l_{radio}$ | 2.00 | 2.00 | 2.77 | 4.55 | 2.00 | 2.00 | 6.07 | 2.77 | 4.55 |
| $l_{TN}$ | 3.192 | - | 10.279 | - | 0.949 | 0.972 | 1.364 | 1.304 | 1.398 |
| $l_{CN}$ | 0.001 | 0.001 | 0.001 | 0.016 | 0.0001 | 0.0001 | 0.002 | 0.0001 | 0.002 |
| $l_{AS}$ | 0.6977 |  | 0.7499 |  | 0.6977 |  |  | 0.7499 |  |
| $l_{E2E}^{singleMNO}$ | 5.897 | - | 13.80 | - | 3.653 | 3.671 | 8.127 | 4.824 | 6.697 |
| $l_{pp}$ | 0.431 |  | 1.493 |  | 0.431 |  |  | 1.493 |  |
| $l_{E2E}^{multiMNO}$ | 6.327 | - | 15.293 | - | 4.083 | 4.102 | 8.558 | 6.317 | 8.190 |

### C. MEC@CN

Table XIII reports the E2E latency values for the MEC@CN network deployment. This deployment requires the V2X traffic to travel through the TN to reach the CN. The larger distances that packets travel through the network and the higher number of packets that are processed by the TN and CN's nodes result in increased values of $l_{TN}$ and $l_{CN}$ compared with the MEC@gNB (Table XI) and MEC@M1 (Table XII) deployments. This challenges this deployment to satisfy the V2X latency requirements. For example, Table XIII shows that this deployment cannot satisfy any of the considered V2X services even under the lowest traffic load when $\alpha$ =0.001. $\alpha$ needs to be increased to 0.0014 and 0.0016 for the LLoA and HLoA V2X services, respectively, under the single MNO scenario and the lowest traffic load ($\lambda_{gNB}^{UL}$ = 2080 pkts/s). These values increase to 0.0055 and 0.0061 when the load increases to $\lambda_{gNB}^{UL}$ = 8320 pkts/s. $\alpha_{min}$ also increases under multi-MNO scenarios. For example, $\alpha_{min}$ values of 0.0021 and 0.0074 are necessary for the HLoA V2X services under 2080 pkts/s and 8320 pkts/s. We should note that under the highest load ($\lambda_{gNB}^{UL}$ = 31200 pkts/s), the MEC@CN deployment cannot satisfy the latency requirements even when $\alpha$=0.01. In this case, $\alpha_{min}$ needs to increase to 0.021 for the LLoA V2X service.

TABLE XIII. E2E LATENCY (IN MS) FOR THE MEC@CN DEPLOYMENT.

|  | $\alpha = 0.001$ |  |  |  | $\alpha = 0.01$ |  |  |  |  |
|---|---|---|---|---|---|---|---|---|---|
|  | LLoA |  | HLoA |  | LLoA |  |  | HLoA |  |
| $\lambda_{gNB}^{UL}$ | 2080 | 8320 | 2080 | 8320 | 2080 | 8320 | 31200 | 2080 | 8320 |
| $l_{radio}$ | 2.00 | 2.00 | 2.77 | 4.55 | 2.00 | 2.00 | 6.07 | 2.77 | 4.55 |
| $l_{TN}$ | - | - | - | - | 2.471 | 2.494 | - | 2.833 | 2.928 |
| $l_{CN}$ | - | - | - | - | 0.001 | 0.003 | - | 0.001 | 0.005 |
| $l_{AS}$ | 0.6977 |  | 0.7499 |  | 0.6977 |  |  | 0.7499 |  |
| $l_{E2E}^{singleMNO}$ | - | - | - | - | 5.174 | 5.203 | - | 6.353 | 8.227 |
| $l_{pp}$ | 0.431 |  | 1.493 |  | 0.431 |  |  | 1.493 |  |
| $l_{E2E}^{multiMNO}$ | - | - | - | - | 5.604 | 5.634 | - | 7.846 | 9.720 |

### D. Centralized

Finally, Table XIV shows the E2E latency values for the Centralized deployment. Note that with respect to previous analyses for the MEC-based network deployments, Table XIV includes the $l_{UPF-AS}$ latency that accounts for the Internet latency and that is only present in the Centralized deployment. As shown in Section XII.D, the 90th and 99.99th percentiles of $l_{UPF-AS}$ are equal to 21 ms and 43 ms, respectively. In addition, the Centralized deployment is characterized by a remote peering point in multi-MNO scenarios, which considerably increases $l_{pp}$. As a result, the E2E latency experienced in this deployment do not satisfy the LLoA and HLoA V2X service

TABLE XIV. E2E LATENCY (IN MS) FOR THE CENTRALIZED DEPLOYMENT.

|  | $\alpha = 0.001$ |  |  |  | $\alpha = 0.01$ |  |  |  |  |
|---|---|---|---|---|---|---|---|---|---|
|  | LLoA |  | HLoA |  | LLoA |  |  | HLoA |  |
| $\lambda_{gNB}^{UL}$ | 2080 | 8320 | 2080 | 8320 | 2080 | 8320 | 31200 | 2080 | 8320 |
| $l_{radio}$ | 2.00 | 2.00 | 2.77 | 4.55 | 2.00 | 2.00 | 6.07 | 2.77 | 4.55 |
| $l_{TN}$ | - | - | - | - | 2.471 | 2.494 | - | 2.833 | 2.928 |
| $l_{CN}$ | - | - | - | - | 2.001 | 2.001 | - | 2.001 | 2.001 |
| $l_{UPF-AS}$ | 21 |  | 43 |  | 21 |  |  | 43 |  |
| $l_{AS}$ | 0.6977 |  | 0.7499 |  | 0.6977 |  |  | 0.7499 |  |
| $l_{E2E}^{singleMNO}$ | - | - | - | - | 28.17 | 28.19 | - | 51.35 | 53.23 |
| $l_{pp}$ | 29.867 |  | 99.212 |  | 29.867 |  |  | 99.212 |  |
| $l_{E2E}^{multiMNO}$ | - | - | - | - | 58.04 | 58.06 | - | 150.57 | 152.44 |
13

requirements. Unlike for the MEC-based network deployments, the LLoA and HLoA V2X service requirements cannot be satisfied by increasing $\alpha$ due to the high latency introduced by the Internet and peering points. The E2E latency results reported in Table XIV are in line with the empirical measurements repoted in [12]-[14] for 5G V2N2V connections that utilize a V2X AS located in the cloud; limited details about the network deployments and configurations are provided.

## XV. A Word of Caution about Average Latencies

The previous section has analyzed the capacity of different 5G network deployments to satisfy the requirements of the LLoA and HLoA V2X services defined by the 3GPP (Table III). The 3GPP establishes strict reliability and latency requirements based on the E2E latency that should be satisfied by the 90[th] and 99.99[th] percentiles of the V2N2V transmissions. This section complements the conducted study with the analysis of the average E2E latency - the metric evaluated in most of the existing studies (e.g. [4]-[10]), and compares the conclusions derived when considering average or percentiles of the E2E V2N2V latency.

Table XV reports the average UL+DL radio latency as a function of the traffic load $\lambda_{gNB}^{UL}$; the 90[th] and 99.99[th] percentiles of the radio latency were reported in Table IV. The comparison of Table XV and Table IV already shows how average latency values might provide misleading conclusions about the capacity of certain network deployments to satisfy certain V2X services. For example, the average radio latency is 1.67 ms when $\lambda_{gNB}^{UL}$=10400 pkts/s and we use MCS Table 3 to guarantee the reliability requirements of the HLoA V2X service. On the other hand, the 99.99[th] percentile of the radio latency for this network load is 12.27 ms, which is above the latency requirement (10 ms) for HLoA. Similar differences between average and 90[th] percentile latency are also observed for the LLoA V2X service under high network loads. The average, 90[th] and 99[th] percentile latency values for the TN and CN were already analyzed in Section XII.A and XII.C, respectively. The analysis showed that the network dimensioning (i.e., $\alpha$) can impact the variability of the TN and CN latency, and hence the differences observed between average and 90[th] or 99[th] percentile latency values. For example, Table VI shows similar average and 99.99[th] percentile TN latency values (0.835 ms vs. 0.875 ms) for MEC@M1 when $\alpha$=0.1 and $\lambda_{gNB}^{UL}$=2080 pkts/s. However, significant differences appear when $\alpha$ is set to 0.001: the average latency is equal to 1.856 ms while the 99.99[th] percentile of the latency increases to 10.279 ms (i.e., above the 10 ms latency requirement of the HLoA V2X service). Conclusions derived using average or percentile latency values also differ when considering the latency contribution of peering points to interconnect MNOs. Table VIII shows that the average latency contribution of remote peering points (13 ms) is below the 25 ms requirement of the LLoA V2X service. However, the 90[th] percentile latency value increases to 29.87 ms, i.e., above the latency requirement defined by 3GPP for this service. Similar trends are also observed for the Internet latency. Section VIII reports average Internet latencies within the same country of 10.3 ms. The 90[th] and 99.99[th] percentile Internet latency values increase to 21 ms and 43 ms, respectively, since the Internet cannot control the latency.

The differences observed in the average and percentile values of the radio, TN, CN, peering points and Internet latency contributions impact the E2E V2N2V latency and the conclusions regarding the suitability of certain network deployments to support V2X services with strict requirements. For example, the average E2E latency is equal to 3.86 ms and 4.17 ms for single and multi-MNO scenarios, respectively, in MEC@M1 deployments with $\lambda_{gNB}^{UL}$=2080 pkts/s and $\alpha$ = 0.001. Both averages satisfy the 10 ms latency requirement established by 3GPP for the HLoA V2X service. However, Table XII shows that the 99[th] percentile E2E latency of the V2N2V connections is 13.80 ms and 15.29 ms for single- and multi-MNO scenarios, respectively, and hence the MEC@M1 deployment cannot really satisfy the reliability requirement established by 3GPP for the HLoA V2X service. A similar situation occurs with MEC@gNB deployments under the same conditions in the multi-MNO scenario. The average E2E latency is 3.22 ms, while the 99[th] percentile E2E latency value increases to 10.096 ms (see Table XI). Centralized network deployments in multi-MNO scenarios with local peering points can achieve average E2E latencies (17 ms) below the 3GPP requirement for the LLoA V2X service (25ms) under medium loads ($\lambda_{gNB}^{UL}$ = 8320 pkts/s) with $\alpha$=0.01. However, this network deployment cannot satisfy the reliability requirement established by the 3GPP for the LLoA V2X service as the E2E 90[th] percentile latency values (58.06 ms in Table XIV with remote peering point or 28.6 ms with local peering point) are above the latency requirement of 25 ms.

TABLE XV. AVERAGE UL+DL RADIO LATENCY IN MS [16]

| LoA | $\lambda_{gNB}^{UL}$ /pkts/s/ | | | | | | | | | |
|---|---|---|---|---|---|---|---|---|---|---|
| | 1040 | 2080 | 4160 | 5200 | 6240 | 8320 | 10400 | 20800 | 31200 | 41600 |
| Low | 1.50 | 1.50 | 1.50 | 1.50 | 1.50 | 1.50 | 1.50 | 1.50 | 1.56 | 3.09 |
| High | 1.50 | 1.50 | 1.51 | 1.52 | 1.53 | 1.58 | 1.67 | 7.31 | 11.81 | 14.23 |

## XVI. Discussion and Conclusions

The possibility for 5G to support advanced V2X services using V2N or V2N2V communications depends on its capacity to satisfy their latency requirements. Existing related studies mostly focus on evaluating the suitability of the 5G RAN, and are usually constrained by reduced deployments under controlled environments that do not capture all the aspects influencing the E2E latency. However, the network deployment and the location of the V2X AS strongly influences the 5G E2E latency. This study advances the state-of-the-art with a unique and novel 5G E2E latency model for V2N and V2N2V communications. The model considers the latency introduced at the radio, transport and core networks, the Internet, the peering points, and the V2X AS for single-MNO and multi-MNO scenarios and for different 5G network deployments that vary the location of the V2X AS from the edge of the network to the cloud. All network deployments analyzed have been proposed or are supported by related international standardization bodies or industry groups. The latency models presented in this paper and [16] represent a valuable tool for the community to study and dimension 5G network deployments to support V2X services. This study would entail a higher cost if it were done through field trials under different deployments and configurations, and a higher computational cost if it is done through network simulators (in addition to the challenge to accurately implement all the network components). The similarity between the results obtained with our 5G latency models and experimental results demonstrate the validity of our model and its high value to the community that can now derive 5G E2E latency values for different 5G network deployments and configurations without having to deploy complete 5G networks.

The derived models have been utilized to evaluate the E2E latency achieved by different 5G network deployments and configurations when supporting V2X services with different requirements using V2N2V connections. The conducted latency analysis has shown that traditional centralized network deployments that locate the V2X AS at the cloud significantly increase the E2E latency due to the latency introduced by the Internet, and cannot satisfy the latency requirements of the high and low level of automation V2X services under study. 5G network deployments with a MEC installed at the CN, TN or gNB decrease the E2E latency with higher reductions as the MEC is deployed closer to the edge. However, the E2E latency that these deployments can achieve is also influenced by the radio, TN, CN and V2X AS configurations and the V2X traffic load to be supported. This study has considered a single but common configuration of the radio interface (bandwidth of 20 MHz and SCS 30 kHz) that resulted in low radio latency values under light to moderate traffic loads. The study in [16] shows that the radio latency can be reduced using higher SCS or mini-slot transmissions. However, this requires larger bandwidth when the traffic load increases. The latency contributions of the transport and core networks highly depend on the network deployment and the ratio of the allocated link capacities to support the V2X traffic. The paper presents a framework designed to derive the minimum ratio of the link capacities at the transport and core networks that is needed to support the requirements of the V2X services. The obtained results have shown that as the MEC is deployed closer to the core network, higher ratios are needed. This challenges the capacity of 5G to simultaneously support a variety of (V2X and non-V2X) services. The obtained results also show that the MEC nodes require significant power and processing capabilities as they are deployed closer to core network to scale and be able to adequately process the V2X traffic. Finally, the conducted study has shown how evaluating the average performance could provide misleading conclusions on the capability of certain 5G network deployments to satisfy stringent requirements of critical V2X services.